# Naive Quantum Gravity

Roderick Sutherland

Centre for Time, University of Sydney, NSW 2006 Australia

rod.sutherland@sydney.edu.au

A possible alternative route to a quantum theory of gravity is described. The usual path quantises the gravitational field in order to introduce the statistical structure characteristic of quantum mechanics. The procedure followed here instead is to remove the probabilistic element of quantum theory by using the less-common technique of introducing final boundary conditions as well as initial on the source. The relevant quantum formalism then becomes compatible with the non-statistical nature of general relativity and a viable theory can be constructed without difficulty. The view taken here is that the lack of both experimental guidance and theoretical consensus lends justification to considering such less mainstream ideas.

## 1. Introduction

This paper outlines an alternative approach to quantum gravity. It seeks to bypass many of the present obstacles to such a theory by relaxing an implicit assumption which is usually taken for granted, namely that only initial boundary conditions should be involved in the quantum description. Given the large amount of effort being expended on other, more sophisticated approaches to quantum gravity, it is important to recognise that simpler models such as the present one are also available.

The model discussed here employs the usual picture of general relativity wherein all of spacetime exists together in the form of a block universe, with the time dimension treated on a similar footing to the spatial dimensions. In such a picture, including final boundary conditions as well as initial ones can be seen as a natural and perhaps more symmetric possibility, as will be described in Sec. 3. This extra restriction allows an alternative approach to quantising gravity. Instead of starting with the statistical nature of quantum mechanics and therefore attempting to make the gravitational field (or curvature) “statistical” as well via a superposition of states, extra information is introduced via a final boundary condition on the source to remove the inherent probabilities from the quantum mechanical description so that it is compatible with the original, classical form of general relativity. This approach also avoids the basic difficulty of trying to formulate a quantum theory of gravity without a pre-existing spacetime background.

As in standard quantum mechanics, the Born rule will be assumed here as a separate postulate describing the distribution of final states. The discussion will mainly be limited to the case where the various source particles involved are not undergoing any non-gravitational interactions and so are not entangled. A wavefunction description will be employed here, rather than the formalism of quantum field theory, because the issues to be addressed are clearer in this format. In particular, this approach has the advantage of setting aside any ambiguity in the mathematical definition of the energy-momentum tensor and issues of regularisation do not arise [1].

An obvious additional problem which exists in the many-particle case of quantum mechanics is that the description resides in configuration space, whereas a description in 4-dimensional spacetime is needed. This issue is addressed in Sec.7 by demonstrating how the required source term in 4 dimensions can be obtained from a many-particle wavefunction. Finally, it is shown how the inclusion of a final boundary condition also allows the measurement process to be treated as taking place in a continuous manner without any discontinuous collapses.

**2. Field Equation**

In the classical theory of general relativity, the field equation has the familiar form:

$$G^{\mu\nu} = 8\pi T^{\mu\nu} \qquad (\mu,\nu = 0,1,2,3) \qquad (1)$$

where $G^{\mu\nu}$ is the Einstein curvature tensor and $T^{\mu\nu}$ is the energy-momentum tensor. The left hand side of this equation describes the curvature of spacetime resulting in the observed gravitational effects, whilst the right hand side describes the source inducing this curvature. The theory also contains the geodesic equation, which determines the path that matter will follow through the curved spacetime. This second aspect will be addressed in Sec. 5.

There is a need to replace the classical field equation (1) with a quantum mechanical version. Consider first the simplified picture where the source consists of a single particle. A well-known initial step [2] is to formulate a semi-classical theory by replacing the tensor $T^{\mu\nu}$ in Eq. (1) with its simplest counterpart in quantum mechanics, namely the following expression[1]:

$$T^{\mu\nu}(x) = \psi^*(x)\hat{T}^{\mu\nu}\psi(x) \qquad (2)$$

which relates to the expectation value of $T^{\mu\nu}$. Here, $\hat{T}^{\mu\nu}$ and $\psi(x)$ are the energy-momentum operator and the particle's wavefunction, respectively, corresponding to the particular wave equation under consideration, e.g, the Dirac equation generalised to a form applicable in Riemannian curved spacetime[2]. The notation $x \equiv (x^0, x^1, x^2, x^3) \equiv (t, \mathbf{x})$ has been used to represent spatial position $\mathbf{x}$ at time t and highlights the fact that expression (2) is a density quantity as a function of position $\mathbf{x}$. Integrating its components over all space would yield expectation values describing the mean *total* amounts of energy, momentum, etc.

---

[1] See, e.g., [3].

[2] The expression for $T^{\mu\nu}(x)$ given by Eq. (2) is guaranteed by each wave equation to have zero divergence. It is also real in each case, this being possible due to half the partial derivatives involved in the operator $\hat{T}^{\mu\nu}$ acting to the left and half to the right.

Using Eq. (2), the field equation (1) becomes:

$$G^{\mu\nu}(x) = 8\pi\, \psi^*(x)\hat{T}^{\mu\nu}\psi(x) \qquad (3)$$

An obviously unsatisfactory feature of this equation is that the right hand side is "statistical", whereas the left hand side is not. In this regard, $\psi(x)$ must be updated discontinuously after a position measurement and any spatial regions of the wavefunction excluded by the measurement result must instantaneously cease contributing to $T^{\mu\nu}(x)$. This update is meant to occur simultaneously throughout space, which is not possible to define unambiguously while maintaining relativistic covariance since it requires favouring one particular space-like hyperplane in spacetime. The classical/quantum mismatch of Eq. (3) is therefore only acceptable for a semi-classical theory, not full quantum gravity.

It will now be demonstrated how this problem can be avoided by introducing a final boundary condition.

## 3. Final boundary condition

The wavefunction $\psi(x)$ has undergone continuous evolution up to the time t at which $T^{\mu\nu}$ is to be evaluated and it essentially summarises the particle's relevant past up to that time. It can therefore be thought of as the initial boundary condition imposed at time t and will be denoted by the letter i. The way to proceed here will now be made clearer by switching to Dirac notation. Writing the wavefunction at position $\mathbf{x}$ at time t in the form $\langle t,\mathbf{x}|i\rangle$, Eq. (2) becomes:

$$T^{\mu\nu}(t,\mathbf{x}) = \langle i|t,\mathbf{x}\rangle \hat{T}^{\mu\nu} \langle t,\mathbf{x}|i\rangle \qquad (4)$$

Looking at the form of this equation, which contains the initial state in the form of both a bra vector $\langle i|$ and a ket vector $|i\rangle$, the obvious way to include a final boundary condition is to employ a familiar technique [5, 8] and replace one of these initial vectors[3] with a final state vector $|f\rangle$ who's precise nature will be discussed below. A normalising factor $\langle f|i\rangle$ also needs to be included, yielding:

$$T^{\mu\nu}(\mathbf{x};t) = \frac{\langle f|t,\mathbf{x}\rangle \hat{T}^{\mu\nu} \langle t,\mathbf{x}|i\rangle}{\langle f|i\rangle} \qquad (5)$$

Eq. (5) is similar to the well-known weak-value expressions [6, 7], as will be discussed in more detail below.

[3] The lack of symmetry between the initial and final states introduced here by choosing one vector rather than the other to be replaced will be rectified in Eq. (6).

As already pointed out, the initial state $|i\rangle$ at time t will be taken to summarise the particle's relevant past. By symmetry, the interpretation of the final state $|f\rangle$ at time t is assumed here to be similar, except that $|f\rangle$ is evolved backwards in time and will summarise the particle's relevant future. This definition of $|f\rangle$ is in precise analogy to the role played by $|i\rangle$. Note that $|f\rangle$ is independent of $|i\rangle$ and should not be confused with the result of deterministically evolving the latter forwards in time. The term $\langle f|t,\mathbf{x}\rangle$ in Eq. (5) corresponds to a separate "final wavefunction" which is taken as having been evolved back to the time at which $T^{\mu\nu}$ is to be evaluated. Both wavefunctions $\langle t,\mathbf{x}|i\rangle$ and $\langle t,\mathbf{x}|f\rangle$ are assumed to satisfy the same wave equation.

The difference in the meaning and behaviour of the two wavefunctions can be appreciated further by considering the simple example of a particle propagating away from a source and towards a detector. The particle's initial wavefunction typically spreads out forwards in time as we evolve it mathematically from the source towards the detector. Its final wavefunction, on the other hand, typically spreads out backwards in time as we evolve it from the detection point back towards the source. Equivalently, taking a forwards-in-time perspective, it would be seen to contract gradually towards the detector.

The initial and final boundary conditions can simply be stated at time t without the need to spell out the particle's various experiences through earlier and later times. For simplicity, $|i\rangle$ and $|f\rangle$ can also be discussed in terms of being the results of measurements performed on the particle at an earlier time $t_i$ and a later time $t_f$, respectively. Such a designation of $|i\rangle$ and $|f\rangle$ as measurement outcomes does, however, entail the presence of retrocausal effects, whereby the experimenter's choice of the next observable to be measured has an effect on the prior situation as well as the subsequent one. The possibility that this occurs in quantum mechanics has been discussed by a number of authors[4], since the uncertainty principle leaves room for it to take place undetected and its presence would resolve certain interpretational difficulties.

It should be noted that the reference to measurements here is not intended to imply that measurement interactions have any special status compared with other interactions, but merely to simplify the presentation. The analysis of the measurement process presented in Sec. 8 indicates that the inclusion of a final boundary condition can enable the process to be viewed as taking place in a continuous manner without any discontinuous changes.

As previously mentioned, Eq. (5) resembles the familiar expressions for weak values. To be precise, it is a density quantity dependent on position $\mathbf{x}$ and integrating it over all space would yield weak values corresponding to each component of the operator $\hat{T}^{\mu\nu}$, in accordance with the standard weak-measurement formalism. Aside from the dependence on position $\mathbf{x}$, the description given by expression (5) may be different from the usual weak-value concept in other ways. A weak value can be given the cautious interpretation of describing the average of a spread of random outcomes which arise from a series of weak measurements on an ensemble and which are not indicative of any particular underlying value. Here, however, the stronger assumption will be made that the calculated value is actually the precise (but hidden) value of the quantity in question, with the spread of values found upon measurement being simply due to the necessarily weak nature of the measurements. This assumption has also been applied in other contexts [5, 17]. Another difference is that weak values are usually discussed in the context of the Two-State Vector Formalism [19], where pre-eminence in terms of elements of reality is usually given to the state vectors, rather than to the weak values. The latter are then viewed as a consequence of the state vectors and are assumed to arise at the time of the weak observation, in accordance with the orthodox quantum mechanical picture. Here, by contrast, they will be taken as fundamental properties existing continuously through time.

[4] For a sample of the large literature on retrocausality in quantum mechanics, see e.g., [4, 5, 8-18].

The intention here is for the new expression (5) to give a more specific valuation for $T^{\mu\nu}(\mathbf{x};t)$ as a result of including the final boundary condition. A preliminary adjustment is needed here to make the new expression consistent with spacetime being real and to make it fully symmetric with respect to the states $|i\rangle$ and $|f\rangle$. Both of these requirements can be achieved via the single step of taking the real part of Eq. (5). This recourse, together with re-employing the abbreviation $x \equiv (t,\mathbf{x})$, then yields[5]:

$$T^{\mu\nu}(x) = \mathrm{Re}\frac{\langle f|x\rangle \hat{T}^{\mu\nu}\langle x|i\rangle}{\langle f|i\rangle} \tag{6}$$

Thus Eq. (6) is assumed here to represent what actually exists in terms of energy and momentum densities at the time where the two boundary conditions are specified. With each of the quantum mechanical wave equations it is easily shown that expression (6) has zero divergence, in analogy to the usual expression (2).

## 4. Non-statistical energy-momentum tensor

Since expression (6) is to be viewed as relating to actual values, not mean values, it therefore resolves the statistical/non-statistical mismatch associated with the semi-classical version in Eq. (3). The statistical nature of the standard quantum mechanical expression in Eq. (4) is then assumed to arise from our lack of knowledge of the final boundary condition. In this regard, it is confirmed in Sec. 6 that Eq. (6) reduces back to Eq. (4) in the usual situation

[5] Note that the new expression can be viewed as the conditional value of $T^{\mu\nu}(x)$ given that prior and subsequent measurements have yielded outcomes $|i\rangle$ and $|f\rangle$, respectively. This conditional wording automatically excludes the unwelcome possibility of the denominator in (6) being zero because a zero value for the probability $|\langle f|i\rangle|^2$ would contradict the stipulation that the $|i\rangle$ and $|f\rangle$ outcomes have actually occurred.

where the final condition is unknown and a weighted average needs to be taken. The general approach of removing probabilities from quantum mechanics by introducing a final boundary condition and thereby obtaining definite values can be viewed as providing the basis for developing an alternative interpretation of quantum mechanics, alongside many-worlds, de Broglie-Bohm, etc.

One obvious way of explaining why the new expression for $T^{\mu\nu}(x)$ need no longer be statistical is that it contains the actual outcome $|f\rangle$ of the next measurement and so other possible outcomes and their probabilities have become irrelevant. The above procedure of specifying the final state demonstrates that there is actually no need to relinquish the classical, non-statistical description of spacetime curvature in order to construct a quantum theory of gravity. For the simple case considered so far of a single source particle, a model can be demonstrated using the following form for the field equation:

$$G^{\mu\nu}(x) = 8\pi \, \mathrm{Re} \frac{\langle f|x\rangle \hat{T}^{\mu\nu} \langle x|i\rangle}{\langle f|i\rangle} \tag{7}$$

The generalisation to a many-particle non-entangled source is also straightforward. Each source particle can still be represented using single-particle wavefunctions for its initial and final states and a separate $T^{\mu\nu}$ term can then be calculated for each particle, with the overall source being just the sum of these terms. This total can then be inserted in the classical field equation (1):

$$G^{\mu\nu} = 8\pi T^{\mu\nu}_{total} \tag{8}$$

For the more general case of an entangled source, the fact that the physical reality will remain in the 4 dimensions of spacetime is shown in Sec. 7.

**5. Counterpart to the classical geodesic equation**

An additional feature of the classical theory which needs to be taken into account is the geodesic equation, which describes the motion of a test particle experiencing the local curvature of spacetime. This classical equation is to be replaced in the proposed model by an appropriate relativistic wave equation for the test particle's initial and final wavefunctions $\langle x|i\rangle$ and $\langle x|f\rangle$. For example, in the Dirac case these wavefunctions would satisfy:

$$\left(i\gamma^{\mu}D_{\mu} - m\right)\langle x|i\rangle = 0 \tag{9}$$

$$\left(i\gamma^{\mu}D_{\mu} - m\right)\langle x|f\rangle = 0 \tag{10}$$

Here, the Dirac equation has been generalised by way of the covariant derivative $D_{\nu}$ to a form applicable in Riemannian spacetime. The metric tensor $g^{\mu\nu}$ underlying this generalisation describes the curved geometry in which the test particle's wavefunctions are evolving. Note that no potential term needs to be inserted in the above equations in order to include gravity. Here gravity is geometric and so is different from other interactions.

Speaking generally, Eqs. (9) and (10) describe the influence of the geometry on a particle's wavefunctions, whilst Eqs. (6) and (8) describe the influence of a particle's wavefunctions on the geometry. Within the context of this non-entangled Dirac case, these four equations comprise a basis for developing a theory of quantum gravity.

Some further aspects of this overall approach will now be presented to provide additional clarity.

**6. Consistency with the standard energy-momentum tensor of quantum mechanics**

For the purposes of the short proof in this section, the final state $|f\rangle$ will be taken here as a measurement result for simplicity and so it will correspond to one of the eigenstates of a complete orthonormal set.

If the conditional dependence of the energy-momentum tensor on both $|i\rangle$ and $|f\rangle$ is included explicitly in the notation, the left hand side of Eq. (6) should really be written as $T^{\mu\nu}(x|i,f)$. In contrast, the standard energy-momentum expression (4) predicted by quantum mechanics should be written as $T^{\mu\nu}(x|i)$, since it depends only on the initial state. Using this notation, it will now be confirmed that expression (6) reduces to the standard one in the usual situation where the final state is not known and a weighted average is taken over the possibilities. Under this procedure, the two expressions are expected to be related via the following familiar relationship for conditional probabilities:

$$T^{\mu\nu}(x|i) = \int_{-\infty}^{+\infty} T^{\mu\nu}(x|i,f)\rho(f)\,df \tag{11}$$

where $\rho(f)$ is the probability density for the subsequent measurement result f. Inserting the usual Born expression:

$$\rho(f) = |\langle f|i\rangle|^2 \tag{12}$$

together with (6) into Eq. (11) yields:

$$\begin{aligned} T^{\mu\nu}(x|i) &= \int_{-\infty}^{+\infty} \mathrm{Re}\frac{\langle f|x\rangle \hat{T}^{\mu\nu}\langle x|i\rangle}{\langle f|i\rangle}|\langle f|i\rangle|^2\,df \\ &= \mathrm{Re}\int_{-\infty}^{+\infty}\langle i|f\rangle\langle f|x\rangle\hat{T}^{\mu\nu}\langle x|i\rangle\,df \\ &= \mathrm{Re}\left[\langle i|x\rangle\hat{T}^{\mu\nu}\langle x|i\rangle\right] \end{aligned} \tag{13}$$

This result is then seen to be equivalent to the standard energy-momentum expression (4) as required, since the factor in the square bracket is already real, as noted in footnote 2.

**7. Entangled many-particle source**

It will now be shown how to proceed in the more general case where the source particles have an entangled wavefunction due to mutual, non-gravitational interactions in the past, even though they are not interacting at present. Leaving open the question of what wave equation this wavefunction would satisfy, such a wavefunction would reside in configuration space and would seem not compatible with the desired 4-dimensional description associated with general relativity. To demonstrate that this is not a problem, the generalisation of Eq. (6) for such an entangled source will now be formulated.

As a first step, consider an initial entangled state $\langle \mathbf{x},\mathbf{x}'|\mathrm{i}\rangle$ describing a pair of particles with coordinates $\mathbf{x}$ and $\mathbf{x}'$. Focussing on the first particle, the density of an arbitrary observable Q for this particle is known in standard quantum mechanics to be expressible as[6]:

$$Q_1(\mathbf{x}) = \int \langle \mathrm{i}|\mathbf{x},\mathbf{x}'\rangle \hat{Q}_1 \langle \mathbf{x},\mathbf{x}'|\mathrm{i}\rangle \, d^3x' \tag{14}$$

where the single-particle operator $\hat{Q}_1$ is acting only on the 1st particle (i.e., on coordinates $\mathbf{x}$, not $\mathbf{x}'$) and the 2nd particle's spatial coordinates are integrated out. Note that, although this integral causes a loss of entanglement information there is no problem here, the reason being that the aim is merely to extract values associated with just one particle, not to retain correlations needed for evaluating joint probabilities, etc.

For the case of the energy-momentum tensor in particular, the appropriate formula is:

$$T_1^{\mu\nu}(\mathbf{x}) = \int \langle \mathrm{i}|\mathbf{x},\mathbf{x}'\rangle \hat{T}_1^{\mu\nu} \langle \mathbf{x},\mathbf{x}'|\mathrm{i}\rangle \, d^3x' \tag{15}$$

where this $T^{\mu\nu}$ is seen to reside in spacetime, not configuration space. This expression now needs to be extended to include the final boundary condition. By analogy with the transition from Eq. (4) to Eq. (6), the result is:

$$T_1^{\mu\nu}(\mathbf{x}) = \mathrm{Re}\int \frac{\langle \mathrm{f}|\mathbf{x},\mathbf{x}'\rangle \hat{T}_1^{\mu\nu} \langle \mathbf{x},\mathbf{x}'|\mathrm{i}\rangle}{\langle \mathrm{f}|\mathrm{i}\rangle} \, d^3x' \tag{16}$$

Turning to the 2nd particle, the corresponding expression is:

$$T_2^{\mu\nu}(\mathbf{x}) = \mathrm{Re}\int \frac{\langle \mathrm{f}|\mathbf{x}'',\mathbf{x}\rangle \hat{T}_2^{\mu\nu} \langle \mathbf{x}'',\mathbf{x}|\mathrm{i}\rangle}{\langle \mathrm{f}|\mathrm{i}\rangle} \, d^3x'' \tag{17}$$

---

[6] See, e.g., [20] Eq. (14.14) and [5] Sec. 7.

where the primes have been deleted from the 2nd particle's coordinates for convenience and the single-particle operator $\hat{T}_2^{\mu\nu}$ now acts only on the 2nd particle's coordinates. This formalism can easily be extended by analogy to N particles. The total energy-momentum tensor in spacetime (not configuration space) will then simply be the sum of the terms for the individual particles:

$$T_{total}^{\mu\nu} = T_1^{\mu\nu} + T_2^{\mu\nu} + ... \tag{18}$$

with each term separately having zero divergence. This total $T^{\mu\nu}$ can then be inserted into the field equation (8).

**8. Continuity in the measurement process**

Since the initial and final boundary conditions have been described in some previous sections in terms of measurement outcomes for convenience, it will be explained here how measurements in general within this picture do not involve any discontinuous changes in the energy-momentum tensor. Rather, the measurement process will be seen to take place continuously and smoothly, evolving via the relevant wave equation. For the purposes of this discussion it will be more convenient to talk in terms of wavefunctions again rather than state vectors. Also, the measuring apparatus will be taken as a macroscopic entity which can be treated classically for simplicity. The more general case is discussed in [8]. Finally, the argument will be implicitly framed in terms of a single particle with a discrete spectrum of outcomes, it being easily generalised to the many-particle, continuous spectrum case.

In standard quantum mechanics, we evolve the usual wavefunction forwards in time to the moment at which a measurement is to be performed. This wavefunction will be referred to here as the "initial wavefunction" $\psi_i$. As argued earlier, a separate (and generally different) "final wavefunction" $\psi_f$ can also be introduced and this can be evolved mathematically backwards in time to the moment of measurement. Now an essential feature of any measurement is that it must allow us to distinguish between the possible outcomes and identify the result. As noted by other authors (e.g., [21]), this means that the possible eigenstates of the observed system (or of something with which it interacts) must become separated in space. Hence, in the simplest case, the initial wavefunction $\psi_i$ will gradually spread into a collection of spatially non-overlapping wave packets in passing through the region of the measurement interaction[7]. This will take place by continuous evolution and is the basis of the well-known Bohm theory of measurement [22].

Turning now to the backwards-evolved final wavefunction $\psi_f$, this wavefunction will also split into separate eigenstates in passing through the measurement region, but the splitting will be in the opposite time direction. Perhaps a useful way of viewing the measurement process is to picture it in spacetime, where the initial wavefunction will spread forwards in time like separate fingers from a hand, while the final wavefunction will spread into fingers backwards in time. From a forwards perspective, the final wavefunction will therefore be seen to merge rather than split during the measurement interaction.

[7] More generally, this process would apply to the combined particle/apparatus wavefunction, with the spatial separation being in configuration space, e.g., different pointer readings, but the macroscopic apparatus is being treated as classical here for simplicity.

This allows the following argument. The branches of the initial wavefunction must become permanently separated after the measurement and not recombine, as otherwise the measurement would be “undone”. Hence it can be expected that the final wavefunction will be non-zero in only one of these branches since it does not split forwards in time. In particular, it will be overlapping only that branch of $\psi_i$ which corresponds to the measurement outcome, thereby selecting this result. A key point to note here is that the new energy-momentum tensor (6) essentially contains a product of the initial wavefunction $\psi_i$ and the final wavefunction’s conjugate $\psi_f^*$. This means that it will be non-zero only in regions where both the initial and final wavefunctions are non-zero. Putting this another way, a branch of the initial wavefunction after the measurement will make no contribution to $T^{\mu\nu}$ unless it is at least partially overlapped in spacetime by the final wavefunction. Consequently, even though the non-overlapped branches of $\psi_i$ evolve on continuously after the measurement, they can be ignored as irrelevant because they contain no energy or momentum. Such “deletion by choice” involves no actual physical collapse and so this picture is consistent with a “no-collapse” view of wavefunction evolution.

In conclusion, since both the initial and final wavefunctions evolve continuously and smoothly during the measurement interaction, their overlapping region of non-zero energy and momentum will also be free of discontinuities and propagate continuously and smoothly.

## 9. Discussion and conclusions

A simple quantum gravity model has been presented in which the approach has been to make the energy-momentum tensor $T^{\mu\nu}$ of quantum mechanics become definite, i.e., “non-statistical”, and therefore compatible with the classical expression for the Einstein tensor $G^{\mu\nu}$. The tensors $G^{\mu\nu}$ and $T^{\mu\nu}$ can then be combined consistently in the field equation of general relativity. This is the opposite of the usual approach, which requires gravity to be quantised to make it “statistical” via a superposition of states. One notable consequence of the model suggested here is that gravity can be incorporated into quantum theory without the need to introduce gravitons. This means that the model is potentially testable and falsifiable via experiment. Formulating such a model has been made possible by the inclusion of the less common notion of imposing a final boundary condition on the source. Finally it should be emphasised that, due to the paucity of experimental evidence available in the domain of quantum gravity, a model of this type is just as viable as the more well-known approaches, while being much simpler.